\documentstyle[12pt,a4,epsf]{article}
\begin{document} 
\bibliographystyle{plain} 
\input{epsf} 
{\Large\bf\noindent Metastable states of  spin glasses on random thin graphs} 
\vskip 2 truecm
\noindent D. S. Dean
\vskip 1 truecm
\noindent 
IRSAMC, Laboratoire de Physique Quantique, Universit\'e Paul Sabatier, 
 118 route de Narbonne, 31062 Toulouse Cedex
\pagestyle{empty} 
\vskip 1 truecm \noindent{\bf Abstract:} 
In this paper we calculate the mean number of metastable states 
for spin glasses on so called random thin graphs with couplings taken from 
a symmetric binary distribution $\pm J$. 
Thin graphs are graphs where the local connectivity of each site is fixed to 
some value $c$. As in totally connected mean field models we find that the number of metastable states increases exponentially with the system size.
Furthermore we find that the average number of metastable states 
decreases as $c$  in agreement 
with previous studies showing that finite connectivity corrections of order 
$1/c$ increase the number of metastable states with respect to the totally 
connected mean field limit. We also  prove that the average number of
metastable states in the limit $c\to\infty$ is finite and converges to
the average number of metastable states in the Sherrington-Kirkpatrick model.
An annealed calculation for the number of metastable states $N_{MS}(E)$ 
of energy $E$ is also carried out giving a lower bound on the ground 
state energy of these spin glasses. For small $c$ one may obtain analytic 
expressions for $\langle N_{MS}(E)\rangle $.
other random models. 
\vskip 1 truecm
\noindent  December 1999
\newpage
\pagenumbering{arabic} 
\pagestyle{plain}
\section{Introduction} 
The nature of the spin glass phase in finite dimensions is still, after 
more than twenty years of study, a widely contested area of condensed matter
physics \cite{fihe}. 
One one hand the droplet or scaling picture \cite{fihu,bmscaling}
 suggests that finite 
dimensional spin glasses may be described by a two phase picture as in a 
ferromagnet and on the other hand mean field calculations on spin glass
models suggest that there are an exponentially large number of {\em pure} 
states organised in an ultrametric geometry coming from the Parisi scheme of
replica symmetry breaking \cite{mepavi}.
 If the latter point of view is correct one should
certainly see at zero temperature a large number of metastable states, 
reflecting the complexity of the free energy landscape
(however the reader is referred to the interesting discussion of the relevance of metastable states to pure states in \cite{nest}). There has been a 
considerable amount of effort to analyse the metastable states in the 
Sherrington-Kirkpatrick mean field model 
\cite{taed,new,rob,dean,olfo,olfost}
 and also the number of solutions of the TAP mean 
field equations for this model (the generalisation of metastable 
states to finite temperature) \cite{brmo1,brmo2}. 
Calculations on this model demonstrate the 
existence of an exponentially large (in terms of the number of spins $N$)
number of metastable states and the continuing existence of a macroscopic
entropy of metastable states even at arbitrarily high vales of a uniform 
magnetic field (in agreement with the divergence of the Almeida Thouless
line at zero temperature) \cite{alth}. This latter fact is clearly a pathology of the totally connected geometry of the SK model. In the SK model each spin 
is connected to all the other spins and the existence of the thermodynamic limit is ensured by scaling the couplings by a factor $1/\sqrt{N}$ in the case of
symmetric distributions. This scaling of the interaction strength with the 
system size is clearly undesirable when one wishes to make a connection with 
the finite dimensional analogue, the Edwards Anderson model, and as remarked 
earlier the survival of the Parisi scheme of replica symmetry breaking 
to finite dimensional systems is still hotly debated. Corrections to order
$1/c$ about mean field theory \cite{taed} seem to suggest an enhancement of the number of metastable states when the dimension is reduced - at odds with the 
naive intuition that increasing the connectivity should increase the complexity
of the system and hence give rise to more local minima in the energy landscape.

In this paper we present the calculation of the average number of metastable 
states for spin glasses on thin graphs, 
these models have been extensively studied in \cite{thin} and whilst accessible to mean field treatment each spin interacts with a fixed number 
(denoted here by $c$) of other spins. It is clear that in these models 
a sufficiently large external field will impose the existence of a single 
metastable at zero temperature. In this case the nature of the distribution 
of the interactions may be expected to play a role, in the Sherrington-Kirkpatrick model the nature of the interactions is wiped out by the central 
limit theorem. Another model of finite connectivity but where 
the local connectivity fluctuates is the Viana Bray model, this was the 
first of these type of models to be studied \cite{vibr}. 
The drawback with these types
of finite connectivity models is that the spin glass phase must be 
characterised by all of the multi-spin overlaps possible (in the SK model
one can use simply the two spin overlap $\langle \sigma_a \sigma_b \rangle$
where $a$ and $b$ are distinct replica indices
(see \cite{mon} for a very clear discussion of the replica formalism 
for dilute models). This additional technical 
difficultly to date has hindered the application of a replica symmetry breaking
scheme in these models, although it has been shown that the replica symmetric
solution is neither stable at the transition temperature, nor at zero 
temperature in a number of cases \cite{instab}. 

The advantage of a metastable state calculation is that, while it 
only gives an indication of the possible nature of the spin glass phase,
one may carry out calculations without replicas.
 
\section{Basic formalism}

 The spin glass model we shall consider has the Hamiltonian 
\begin{equation}
H = -{1\over 2} \sum_{j\neq i} J_{ij} n_{ij} S_i S_j
\end{equation} 
 where the $S_i$ are Ising spins, $n_{ij}$ is equal to one if the sites $i$ and $j$ are connected and the $J_{ij}$ are taken from a binary distribution where
$J_{ij}= -1$ with probability half and $  J_{ij}= 1$ with probability half.
The techniques used in this paper can also be used in the case
of other symmetric probability distributions, such as the 
Gaussian distribution, however the advantage with the binary distribution is 
that one may find analytical expressions for certain quantities that would 
require numerical solution in the case of other distributions.
A metastable state is defined to be a configuration where if one changes the 
sign of any given spin the energy does not decrease, for the purposes of 
this paper we shall include the {\em marginal}, case where the energy does not 
change, as being metastable. With this definition number of metastable states is
given by \cite{taed,dean,rob}
\begin{equation}
 N_{MS}  = {\rm Tr} \prod_{i=1}^N \theta\left( \sum_{j\neq i} J_{ij}n_{ij} S_i S_j
\right)
\end{equation} 
The fact that we include the marginal case implies that here $\theta(x)$ the 
Heaviside step function is taken such that $\theta(0) =1$. One may exploit
the parity of the distribution of the $J_{ij}$ by making a gauge transformation
$J_{ij} \to J_{ij}S_iS_j$ \cite{taed,dean} to obtain 
\begin{equation}
 \langle N_{MS} \rangle = 2^N \langle \prod_{i=1}^N \theta\left( \sum_{j\neq i} J_{ij} n_{ij}\right) \rangle
\end{equation}
Here we are obliged to discuss the construction of the thin graphs. One may
generate these graphs by considering planar Feynman diagrams as in \cite{bcp},
however we shall construct them by selecting the graphs of fixed local 
connectivity from a larger ensemble of random graphs, that is the type of  
random graph  found in the Viana Bray spin glass model \cite{vibr}. 
These random graphs are simply constructed as follows: any two points are 
connected with probability $p/N$. Hence $n_{ij}$ is equal to one with probability $p/N$ and zero with probability $1-{p\over N}$. Here $p$ is some arbitrary number of order one and we shall see that the results one obtains are independent of the choice of $p$. If we denote the average on a random graph (with a specified value of $p$) by $\langle \cdot\rangle_p$ then the induced 
average over the subset of thin graphs of connectivity $c$ is given by 
\begin{equation} 
\langle F \rangle = 
{\langle  F \prod_{i=1}^N \delta_{\sum_{i\neq j} n_{ij}\ , c} \rangle_p\over
M(N,c,p)}
\end{equation}
where 
\begin{equation}
M(N,c,p) = \langle \prod_{i=1}^N \delta_{\sum_{i\neq j} n_{ij}\ , c} \rangle_p
\end{equation}
is the average number of thin graphs of connectivity $c$ generated by the 
random graph ensemble for a given $p$.
Expressing the Kronecker delta functions as Fourier integrals one finds:

\begin{eqnarray}
M(N,c,p) &=& \langle {1\over (2\pi)^N} \int_0^{2\pi} \prod_i d\lambda_i 
\exp\left( i\sum_{i\neq j} \lambda_i n_{ij} - ic\sum_i \lambda_i\right)\rangle
\nonumber \\
 &=& {1\over (2\pi)^N} \int_0^{2\pi}\prod_i d\lambda_i \exp(-ic\lambda_i) \prod_{i<j} \left(
1 - {p\over N} + {p\over N} \exp\left( i(\lambda_i + \lambda_j)\right)\right)
\nonumber \\
 &=& {1\over (2\pi)^N} \int_0^{2\pi}\prod_i d\lambda_i \exp(-ic\lambda_i)
\exp\left( - {Np\over 2} + {p\over 2 N} \sum_{ij} \exp\left(i(\lambda_i + \lambda_j)\right) + O(1)\right) \nonumber 
\end{eqnarray}

where we have neglected terms of $O(1)$ in the exponential above. We now carry out a Hubbard-Stratonovich transformation yielding

\begin{eqnarray}
M(N,c,p) &=& {N^{1\over 2} \over (2\pi)^{N + {1\over 2}}}
\int dz \exp\left(- {N z^2\over 2} - 
{Np\over 2} \right )\prod_i \exp\left( -ic\lambda_i 
+ p^{{1\over 2}}z \exp(i\lambda_i) \right) d\lambda_i \nonumber \\
&=& A\int dz\ \exp(NS[z])
\end{eqnarray}
where $A$ is a constant term containing non-extensive terms in $N$ and 

\begin{eqnarray}
S[z] &=& {-z^2\over 2} + \log\left( \int_0^{2\pi} d\lambda \exp\left( -ic\lambda + p^{1\over 2} z \exp(i\lambda)\right)\right) - \log(2\pi) -{p\over 2} 
\nonumber \\
     &=& {-z^2\over 2} + c \log(z) + {c\over 2} \log(p) -\log(c!) - {p\over 2}
\end{eqnarray}

The integral over $z$ may now be evaluated by the saddle point method at the 
maximum of $S$ given by its value at $z^* = c^{1\over 2}$ leading to
the final result

\begin{equation} 
{\log(M(N,c,p))\over N} = \sup_{z} S[z] = 
{1\over 2}c\left( \log(c) + \log(p)  -1\right) -\log(c!) -{p\over 2} \label{eq:nop}
\end{equation}
It is worth noting that the value of $p$ maximising the average number of 
thin graphs of connectivity $c$ is $p = c$ as one would expect.

In the notation set up so far we have that
\begin{equation}
 \langle N_{MS} \rangle  = {D(N,c,p)\over M(N,c,p)}
\end{equation}
where 
\begin{equation}
D(N,c,p) = \langle\langle \ \langle \prod_{i=1}^N \theta\left( \sum_{j\neq i} J_{ij} n_{ij}\right)   \delta_{\sum_{i\neq j} n_{ij}\ , c} \rangle_p  
 \ \rangle \rangle_J
\end{equation}
where $\langle\langle \cdot \rangle \rangle_J$ indicates the average
over the couplings.
Using the representation 
\begin{equation}
\theta(z) = {1\over 2\pi} \int_{0^-}^\infty dx \int_{-\infty}^\infty d\lambda^*
\exp(-i\lambda^* (z-x))
\end{equation}
we obtain
\begin{eqnarray}
D(N,c,p) = \langle\langle  && \langle {1\over 2^N \pi^{2N}}\int \prod_i d\lambda_id\lambda^*_i
dx_i \exp\left( -ic\sum_i \lambda_i +\sum_i x_i\lambda_i^*+ i\sum_{i\neq j} n_{ij}\lambda_i \right. \nonumber  \\
&-& \left.
i \sum_{i<j} n_{ij} J_{ij} (\lambda_i^* + \lambda_j^*)\right)\rangle_p  
\ \rangle
\rangle_J
\end{eqnarray}
Carrying out the average over the $n_{ij}$ one obtains
\begin{eqnarray}
D(N,c,p) = {1\over 2^N \pi^{2N}}&&\int \prod_i d\lambda_i d\lambda^*_i
dx_i \exp\left( -ic\sum_i \lambda_i +\sum_i x_i\lambda_i^* -{Np\over 2}
\right. \nonumber \\ 
+ && \left.{p\over 2N} \sum_{ij} \langle\langle \exp\left(i\left(\lambda_i + \lambda_j
 - J_{ij}(\lambda_i^* + 
\lambda_j^*)\right)\right) \rangle \rangle_J \right)
\end{eqnarray}
We emphasise that the disorder average here is an annealed one as one is 
computing $\langle N_{MS} \rangle $ and not $\langle \log(N_{MS}) \rangle$. 
For the symmetric binary distribution considered here one finds

\begin{eqnarray}
&&D(N,c,p) = {1\over 2^N \pi^{2N}} \int\prod_i d\lambda_i d\lambda^*_i
dx_i \exp\left( -ic\sum_i \lambda_i +\sum_i x_i\lambda_i^* -{Np\over 2}
\right. \nonumber \\ 
&+& \left. {p\over 4N} \sum_{ij} \exp\left(i(\lambda_i + \lambda_j - 
\lambda_i^* - \lambda_j^*)\right) +{p\over 4N} \sum_{ij} \exp\left(i(\lambda_i + \lambda_j + \lambda_i^* + \lambda_j^*)\right) \right) 
\end{eqnarray}

Making a Hubbard Stratonovich transformation to decouple the two interacting terms one obtains one may carry out the $\lambda$, $\lambda^*$ and $x$ 
integrations site by site to obtain 
\begin{equation}
D(N,c,p) = {N\over 2\pi 2^N \pi^{2N}}\int dz_- dz_+ \exp(N S^{**}[z_+,z_-])
\end{equation}
where
\begin{eqnarray}
S^{**}[z_+,z_-] &=& -{z_+^2\over 2} - {z_-^2\over 2} - {p\over 2} 
+\log\left[ \int d\lambda d\lambda^* dx
 \exp\left( ({p\over 2})^{1\over 2} \right.\right. \nonumber \\ 
&& \left.\left. z_+ \exp\left(i(\lambda + \lambda^*)\right) + ({p\over 2})^{1\over 2} 
z_- \exp\left(i(\lambda - \lambda^*)\right) - ic\lambda + i\lambda^*x \right)\right] \label{eq:int}
\end{eqnarray}

Recalling that the $\lambda$ integration is on $(0,2\pi)$, the $\lambda^*$ integration is on $(-\infty, \infty)$ and the $x$ integration is on $[0^-, \infty)$
(because we have chosen to take $\theta(0) = 1$) one may simplify, after some algebra, the integral in the logarithm in (\ref{eq:int}) to obtain
\begin{equation}
S^{**}[z_+,z_-] = -{z_+^2\over 2} -{z_-^2\over 2} - {p\over 2}  + {c\over 2}\left( \log(p) - \log(2) \right) + 2\log(2\pi) + \log\left[ 
\sum_{m\geq {c\over 2}}^c { z_+^{c-m} z_-^m \over m! (c-m)!}\right]
\end{equation}
The remaining integrals may be evaluated by the saddle point method, 
collecting the extensive terms in $N$ and normalising by the term
$M(N,c,p)$ we find the result
\begin{equation}
{\log(\langle  N_{MS}\rangle )\over N} = \sup_{\{z_+,z_-\}} S^*\left[z_+,z_-\right]
\end{equation}
where
\begin{equation}
S^*[z_+,z_-] = -{1\over 2}z_+^2 - {1\over 2} z_-^2 + \log\left[ 
\sum_{m\geq {c\over 2}}^c {c! z_+^{c-m} z_-^m \over m! (c-m)!}\right]
+ \log(2)(1- {c\over 2}) + {1\over 2}c(1 - \log(c)) \label{eq:main}
\end{equation}
where we see that the dependence on $p$ has disappeared (this is normal as the introduction of $p$ was as a mathematical 
artifact to construct the thin graphs).
Making the change of variables $z^- = \mu z^+$ in equation (\ref{eq:main}) 
one may solve the saddle point equation explicitly for $z^+$ and one finds that 

\begin{equation}
{\log(\langle  N_{MS}\rangle )\over N} = \sup_{\mu} S\left[ \mu \right]
\end{equation}
where 
\begin{equation}
S[\mu]  = -{1\over 2} c \log( 1 + \mu^2)  + \log(2)( 1 - {c\over 2})
+ \log\left[\sum_{m\geq {c\over 2}}^c {c ! \mu^m \over m! (c-m)!}\right]
\end{equation}
\section{Specific examples}
The case $c = 1$ can be easily solved, one finds that
\begin{equation}  
{\log(\langle N_{MS}\rangle)\over N} = {1\over 2}\log(2)
\end{equation}

This result is easy to understand as the spins form dimers, each dimer has
two possible states corresponding to a change in the sign of each of the 
two spins concerned.

The case $c = 2$ is of interest as when $c=2$ the thin graphs correspond
to a collection of large closed chains of spins. Here we find that

\begin{equation}  
{\log(\langle N_{MS})\over N} = \log({1 + \sqrt{5}\over 2}) \label{eq:1d}
\end{equation}

The number of metastable states in one dimensional spin glasses has 
been studied in the case of a continuous  even probability distribution 
$p(J)$ for the $J_{ij}$ by Derrida and Gardner \cite{dega}
and by Li \cite{li}. In this case 
one finds the  results ${\log(\langle N_{MS}\rangle)\over N} = \log({4\over \pi})$ 
and ${\langle\log(N_{MS})\rangle\over N}= \log(2)/3$,  that is these two averages are 
independent of the precise form of $p(J)$. In the case of the binary distribution studied here one may carry out the following extremely easy calculation
\cite{li}.
Using the notation developed earlier one may write in one dimension
\begin{equation}
\langle  N_{MS} \rangle = 2^N \langle \prod_{i=1}^N 
\theta\left(J_{i-1,i} + J_{i,i+1})\right) \rangle
\end{equation}

Define by $Q^+(N)$/$Q^-(N)$ the average number of 
metastable states of a one dimensional spin glass with  $N$ bonds where the 
last ($N$th) bond is taken to be positive/negative. 
The boundary conditions at the end of the chains are taken to be free but this does not change the result in the thermodynamic limit. By recurrence it is 
easy to see that

\begin{eqnarray}
Q^+(N) &=& {1\over 2} Q^+(N-1) + {1\over 2} Q^-(N-1) \\
Q^-(N) &=& {1\over 2} Q^+(N-1)
\end{eqnarray}

Solving these equations we find in the thermodynamic limit the result 
(\ref{eq:1d}). Of course this result is not surprising as it is clear that
the thin graphs generated in the case $c=2$ will generate a number of 
disconnected loops of macroscopic size, the additivity of the entropy of the
metastable states ensures the  equivalence of the two results. What is 
amusing however is that a mean field calculation is capable of reproducing
a transfer matrix calculation for a one dimensional system !

In the  case $c=3$ the saddle point equations remain quadratic and one finds 
that 
\begin{equation}
{\log(\langle N_{MS}\rangle )\over N} = {1\over 2}\log({8\over 5})
\label{eq:c3}
\end{equation}

If one wished to naively mimic a lattice system $c = 3$ could  correspond 
to a honeycomb lattice is two dimensions. This model has been studied numerically by a transfer matrix method in \cite{honey} and in the case of the binary
bond, distribution considered here it was found that $ \log(\langle N_{MS}\rangle )/(N\log(2)) = 0.339$ compared to the prediction of (\ref{eq:c3}) which
gives $0.33903$, leading to the intriguing question whether (\ref{eq:c3})
is an exact result. Even if not an exact result, the random graph approximation
for  the honeycomb lattice problem provides a remarkably accurate result, 
suggesting that many of the properties of finite dimensional lattice spin 
glasses are dominated by the geometry of their local connectedness and not 
by their global topology.

For values of $c\geq 4$ one may solve the equations numerically and the results
are show in figure (1). Note that the higher of the two curves shown  corresponds to even values of $c$ where $\langle N_{MS}\rangle$ can be expected to be 
large as we have included as metastable states those whose energy is unaltered
or increased when flipping a single spin ({\em i.e.} the marginally metastable
states. The two curves tend to the same limit $0.28743$. This is the 
value obtained in the Sherrington Kirkpatrick model by  Tanaka and Edwards
\cite{taed}.
However the approach to the two limits is different. In the case of $c$ odd
the correction is as $1/c$ as in the case of the SK model, in this case there are no marginal metastable state and in the SK model there are almost surely no
marginal metastable states. In the case $c$ even however there are marginal 
metastable states and the correction to the totally connected states is
found to be numerically $1/c^{1\over 2}$. Here we will demonstrate analytically
the convergence to the SK result.

We may rewrite the action $S[\mu]$ as 

\begin{equation}
S[\mu]  = -{1\over 2} c \log( 1 + \mu^2)  + \log(2)( 1 - {c\over 2})
+ + c\log( 1 + \mu) + \log\left[F(\mu)\right]
\end{equation}

where 

\begin{equation}
F(\mu) = \sum_{m\geq {c\over 2}}^c {c ! p^m q^{c-m}\over m! (c-m)!}
\end{equation}
where $p = {\mu\over 1+\mu}$ and $q = 1-p$ may be interpreted as the probability of success and failure of a Bernoulli process. 
Using the central limit theorem one may write in the limit of large $c$ that
$F[\mu] \to P(X \geq {1\over 2})$ where $X$ is a Gaussian random variable
of mean $p$ and variance $pq/c$ (we shall see a postiori that the ansatz on 
the asymptotic behaviour of $\mu$ at its saddle point value and the central
limit theorem approximation are consistent to the same order of approximation).
We now make the ansatz that for $c$ large the saddle point value of $\mu$ has the form $\mu = 1 + a/c^{1\over 2} +\ \ {\rm lower\  order\  terms}$. Using the 
central limit theorem we find
\begin{eqnarray}
F(\mu) &=& \sqrt{{c\over 2 \pi pq}}\int_{1\over 2}^\infty \exp\left(
-{(z-p)^2 c\over 2pq}\right) \nonumber \\
       &=& \sqrt{{1\over  2 \pi }}\int_{{1\over 2} - \sqrt{{cp\over q}}}^\infty \exp\left(-{1\over2}z^2 \right) dz
\end{eqnarray}

The ansatz above on $\mu$ gives $p \sim {1\over 2} + {a\over 4c^{1\over 2}}$ giving 
finally

\begin{equation}
F(\mu) = \sqrt{{1\over  2 \pi }}\int_{-{a\over 2}}^\infty 
\exp\left(-{1\over2}z^2 \right) dz
\end{equation} 
Making the same ansatz throughout the action and developing to leading order
gives 

\begin{equation} 
S[a] = \log(2) -{a^2\over 8} + \log\left[ \sqrt{{1\over  2 \pi }}\int_{-{a\over 2}}^\infty \exp\left(-{1\over2}z^2 \right) dz \right]
\end{equation}

Which is exactly the same as the variational equation in \cite{taed}
leading to the result for the SK model.

\section{Metastable states of fixed energy}

Here we shall examine the average of the number of metastable states  
$N_{MS}(E)$ of with a  given energy $E$ per spin. To exactly calculate
the zero temperature thermodynamic properties of the system on should calculate
the average value of the logarithm of this number, which is probably not
the log of the average value of $N_{MS}(E)$ due to correlations between 
states of the same fixed energy (for example see the discussions in 
\cite{taed, rob}), from Jensen's inequality however one has the bound
$ \langle \log\left( N_{MS}(E) \right) \rangle  
\leq \log\left(\langle N_{MS}(E)\rangle  \right)$. We may write $N_{MS}(E)$
as 
\begin{equation}
 N_{MS}(E)  
= {\rm Tr} \prod_{i=1}^N \delta\left( H - EN \right)
\theta\left( \sum_{j\neq i} J_{ij}n_{ij} S_i S_j
\right)
\end{equation}
The calculation now includes an additional integration to enforce the energy
constraint, which amounts to adding an external uniform field to the fields 
$\lambda_i^*$. One finds that 

\begin{equation}
{\log(\langle  N_{MS}(E)\rangle )\over N} = 
\sup_{\{z_+,z_-, \alpha \}} S^*\left[z_+,z_-, \alpha\right]
\end{equation}
where $\alpha$ is an  additional Lagrange multiplier enforcing the 
constraint on the energy and 
\begin{equation}
S^{*}[z_+,z_-,\alpha ] = -{z_+^2\over 2} -{z_-^2\over 2} + \log\left[ 
\sum_{m\geq {c\over 2}}^c { z_+^{c-m} z_-^m e^{-m\alpha}\over m! (c-m)!}\right]
+ \log(2)\left( 1 - {c\over 2}\right) + {1\over 2} c\left(1 - \log(2)\right)
+\alpha\left( {c\over 2} - E\right)
\end{equation}
making the substitution  $z_- = \mu z_+$, one may then solve the stationarity
equation for $z_+$ as before to obtain 
\begin{equation}
{\log(\langle  N_{MS}(E)\rangle )\over N} = 
\sup_{\{\mu, \alpha \}} S\left[\mu, \alpha \right]
\end{equation}
where
\begin{equation}
S\left[\mu, \alpha \right] = -{1\over 2} c \log( 1 + \mu^2)  + \log(2)( 1 - {c\over 2})
+ \log\left[\sum_{m\geq {c\over 2}}^c {c ! \mu^m e^{-\alpha m}\over m! (c-m)!}\right] + \alpha\left( {c\over 2} - E\right) \label{eq:en}
\end{equation}

For $c\geq 4$ the equations maximising (\ref{eq:en}) may now 
be solved numerically by fixing $\alpha$ and maximising
over $\mu$. The corresponding value for the energy is then given by
\begin{equation}
E = {c\over 2}\left( {1- \mu_*^2( \alpha) \over 1+  \mu_*^2(\alpha) }\right)
\end{equation}

where $\mu_*(\alpha)$ is the value of $\mu$ is that which maximises 
(\ref{eq:en}) for fixed $\alpha$. 

In the cases $c = 2$ and $c = 3$ an analytic solution is possible; one finds
that for $c = 2$ the support of $\langle  N_{MS}(E)\rangle$ is $(-1,0)$ where
\begin{equation}
{\log(\langle  N_{MS}(E)\rangle )\over N}
= {(1-E)\over 2} \log(1-E) - {(1+E)\over 2} \log(1+E) + E\log(2) + E\log(-E)
\end{equation}
In the case $c=3$ the support of $\langle  N_{MS}(E)\rangle$ is $(-{3\over 2}
,-{1\over 2})$ where
\begin{eqnarray}
{\log(\langle  N_{MS}(E)\rangle )\over N} = 
-\log(2) &-{(3+2E)\over 4}\log(3+2E) + {(3-2E)\over 4}\log(3-2E) \nonumber\\
&+ E\log(3) + {(1+2E) \over 2}\log(-1 - 2E)
\end{eqnarray}
Hence of the advantages of this model is that one can have exact analytical
expressions for the density of metastable states in terms of energy (in 
totally connected mean field models only a numerical solution is possible).

Defining 
\begin{equation} 
E^* = \inf \{ E\ : \  {\log(\langle  N_{MS}(E)\rangle )\over N} = 0\}
\end{equation}
from Jensen's inequality it is clear that the ground state energy of the
system $E_g$ is bounded from below by $E^*$, {\em i.e.} $E_g \geq E^*$. This 
value of $E^*$ is shown in figure (3) for values of $c$ up to 30. 

\section{Conclusion}
In conclusion we have seen that in agreement with calculations to order $1/c$, 
where $c$ is the lattice connectivity for the Edwards Anderson model, about
the mean field Edwards Anderson model, decreasing the local connectivity
increases the average metastable states for a spin glass on a thin graph (apart
from the fluctuations that occur on going between even and odd connectivities).
For realistic values of $c$ ({\em i.e.} those that could mimic three dimensional lattice structures) one finds  an exponentially large average number of 
metastable states. Interestingly the calculation for $c=3$ appears to reproduce the numerical calculation of \cite{honey} for the two dimensional 
honeycomb lattice.

\newpage
\begin{figure}[htb]
\begin{center}\leavevmode
\epsfxsize=15 truecm\epsfbox{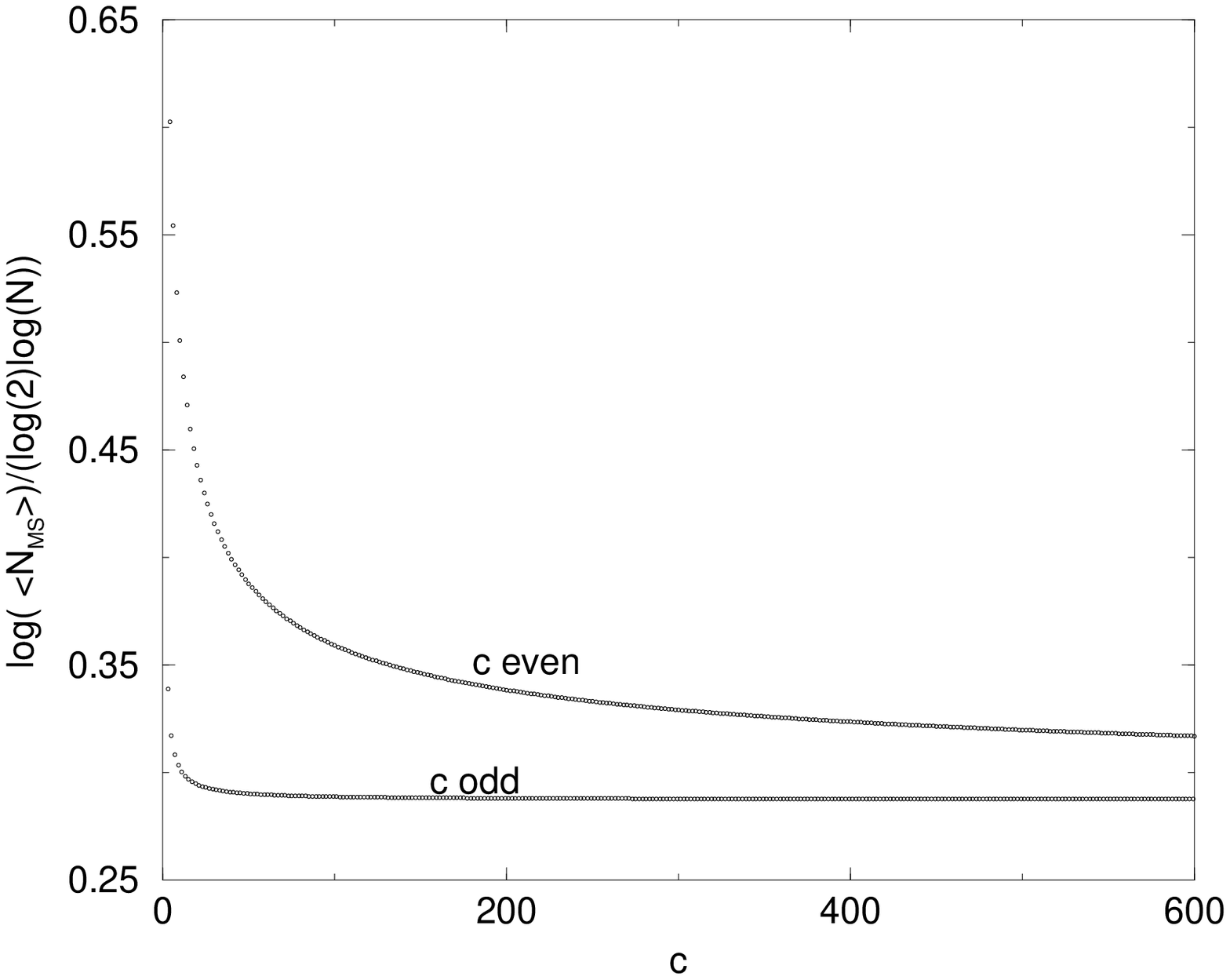}
\end{center}
\caption{${\log(\langle N_{MS}\rangle)\over N\log(2)}$ as a function of $c$ the connectivity} 
\label{figure:Fig1}
\end{figure}
\newpage\begin{figure}[htb]
\begin{center}\leavevmode
\epsfxsize=15 truecm\epsfbox{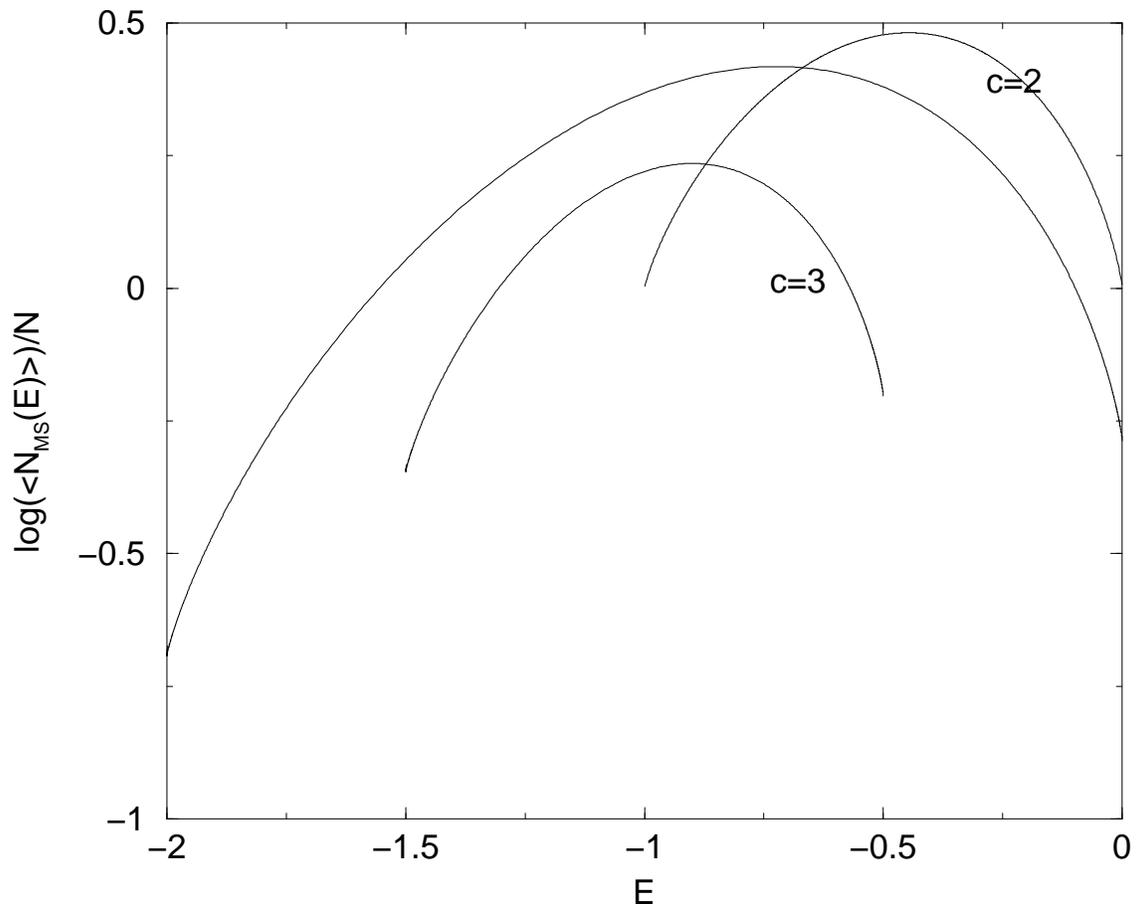}
\end{center}
\caption{${\log(\langle N_{MS}(E)\rangle)\over N}$ for $c=2$, $c=3$ 
and $c=4$}
\label{figure:Fig2}
\end{figure}
\newpage
\begin{figure}[htb]
\begin{center}\leavevmode
\epsfxsize=15 truecm\epsfbox{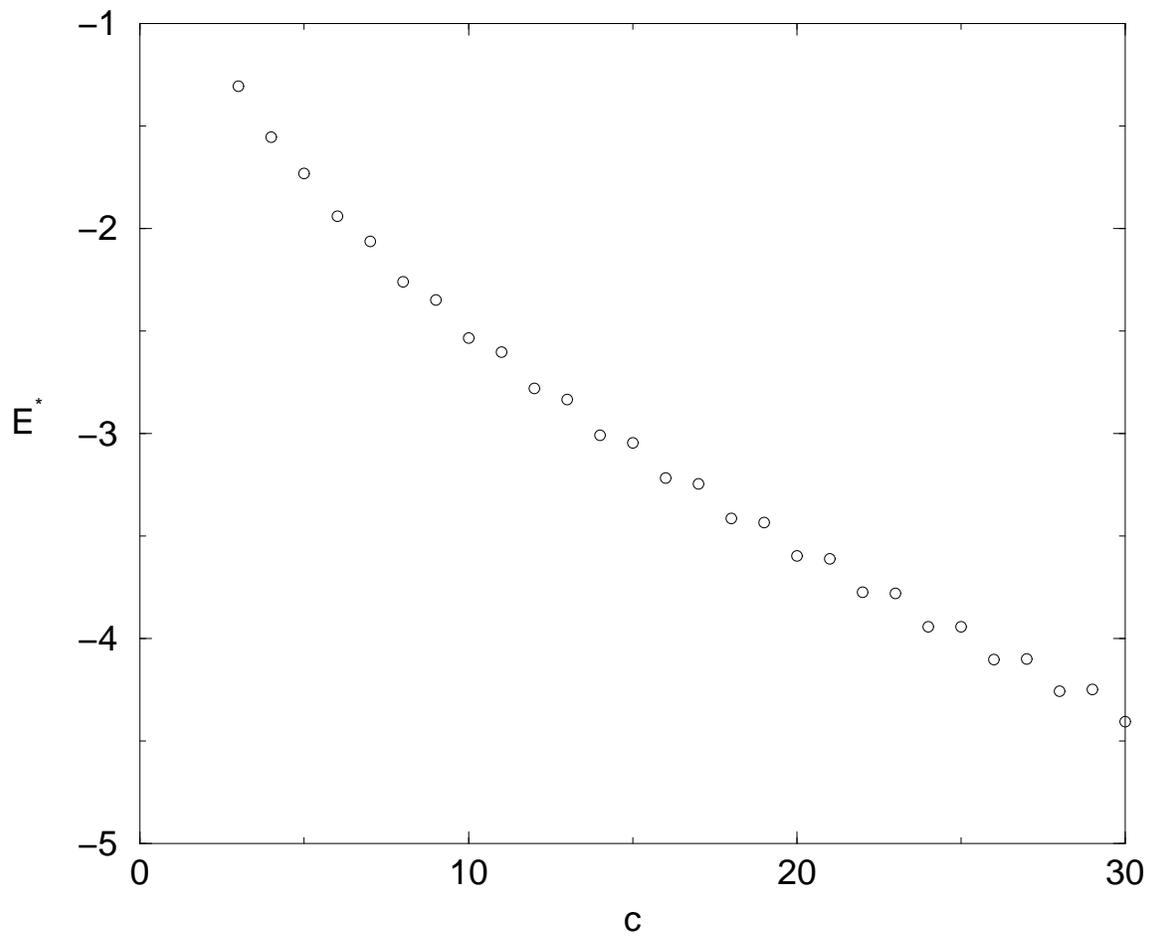}
\end{center}
\caption{Lower bound $E^*$ for the groundstate energy $E_g$ as a function of connectivity $c$} 
\label{figure:Fig3}
\end{figure}
\pagestyle{empty}
\end{document}